# Fire in the sky: The southern lights in Indigenous oral traditions

Duane W. Hamacher

Nura Gili Indigenous Programs Unit, University of New South Wales,
Sydney, NSW, 2052, Australia
Email: d.hamacher@unsw.edu.au

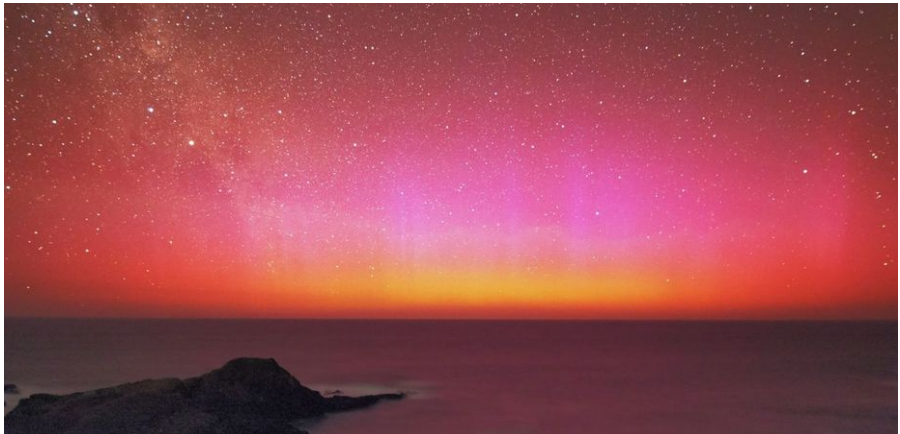

Aurora Australis as seen from Victoria. Alex Cherney, Terrastro Gallery.

Parts of Australia have been privileged to see dazzling lights in the night sky as the Aurora Australis – known as the southern lights – puts on a show this year.

A recent surge in solar activity caused spectacular auroral displays across the world. While common over the polar regions, aurorae are rare over Australia and are typically restricted to far southern regions, such as Tasmania and Victoria.

But recently, aurorae have been visible over the whole southern half of Australia, seen as far north as Uluru and Brisbane.

**Different cultures**

It's a phenomenon that has existed since the Earth's formation and has been witnessed by cultures around the world. These cultures developed their own explanation for the lights in the sky – many of which are strikingly similar.

From a scientific point of view, aurora form when charged particles of solar wind are channelled to the polar regions by Earth's magnetic field. These particles ionize oxygen and nitrogen molecules in the upper atmosphere, creating light.

Auroral displays can show various colours, from white, to yellow, red, green, and blue. They can appear as a nebulous glowing arcs or curtains waving across the sky.

Aurorae are also reported to make strange sounds on rare occasions. Witnesses describe it as a crackling sound, like rustling grass or radio static.

In the Arctic, the Inuit say the noise is made by spirits playing a game or trying to communicate with the living.

In 1851, Aboriginal people near Hobart said an aurora made noise like "people snapping their fingers". The cause of this noise is unknown.

Aurorae are significant in Australian Indigenous astronomical traditions. Aboriginal people associate aurorae with fire, death, blood, and omens, sharing many similarities with Native American communities. They are quite different from Inuit traditions of the Aurora Borealis, which are more festive.

**Fire in the sky**

Aboriginal people commonly saw aurorae as fires in the cosmos. To the Gunditjmara of western Victoria, they're Puae buae ("ashes"). To the Gunai of eastern Victoria, they're bushfires in the spirit world and an omen of a coming catastrophe.

The Dieri and Ngarrindjeri of South Australia see aurora as fires created by sky spirits.

As far north as southwestern Queensland, Aboriginal people saw the phenomenon as "feast fires" of the Oola Pikka —- ghostly beings who spoke to Elders through the aurora.

The Maori of Aotearoa/New Zealand saw aurorae (Tahunui-a-rangi) as the campfires of ancestors reflected in the sky. These ancestors sailed southward in their canoes and settled on a land of ice in the far south.

The southern lights let people know they will one day return. This is similar to an Algonquin story from North America.

**A warning to follow sacred law**

Mungan Ngour, a powerful sky ancestor in Gunai traditions, set rules for male initiation and put his son, Tundun, in charge of the ceremonies. When people leaked secret information about these ceremonies, Mungan cast down a great fire to destroy the Earth. The people saw this as an aurora.

Near Uluru, a group of hunters broke Pitjantjatjara law by killing and cooking a sacred emu. They saw smoke rise to the south, towards the land of Tjura. This was the aurora, viewed as poisonous flames that signalled coming punishment.

The Dieri also believe an aurora is a warning that someone is being punished for breaking traditional laws, which causes great fear. The breaking of traditional laws would result in an armed party coming to kill the lawbreakers when they least expect it.

In this context, fear of an aurora was utilised to control behaviour and social standards.

**Blood in the cosmos**

The red hue of some aurorae is commonly associated with blood and death.

To Aboriginal communities across New South Wales, Victoria, and South Australia, auroral displays represented blood that was shed by warriors fighting a great battle in the sky, or by spirits of the dead rising to the heavens.

Celestial events that appear red are often linked to blood, including meteors and eclipses.

A total lunar eclipse turns the moon red (sometimes called a blood-moon), which was seen by some communities as the spirit of a dead man rising from his grave.

Rare astronomical events were viewed as bad omens by cultures around the world. Now imagine if two of these events overlap!

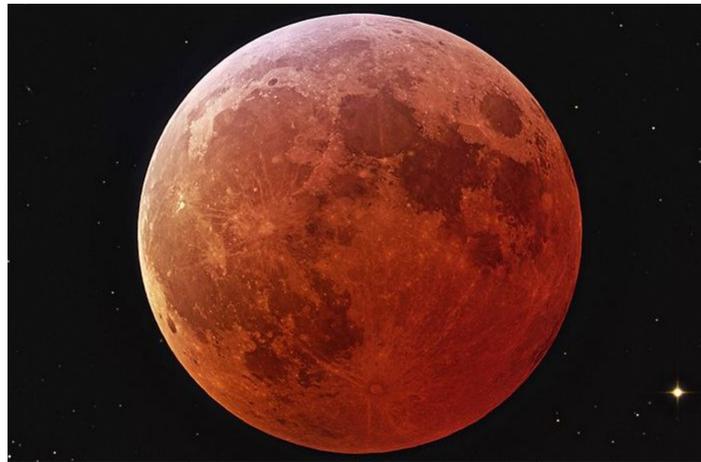

The moon turning red during an eclipse, also known as a blood moon. NASA

In 1859, Aboriginal people in South Australia witnessed an auroral display *and* a total lunar eclipse. This caused great fear an anxiety, signalling the arrival of dangerous spirit beings.

There could be a repeat of this astronomical double-act as a lunar eclipse will be visible across Australia on Saturday April 4, 2015.

Will the aurorae continue? Keep watch.

> *Originally published in The Conversation: https://theconversation.com/fire-in-the-sky-the-southern-lights-in-indigenous-oral-traditions-39113*